\documentclass[12pt]{iopart}
\usepackage[pdftex]{graphicx}
\usepackage{cite}
\usepackage{xspace}
\usepackage{color}
\usepackage{amssymb}
\newcommand{\bite}{Bi$_{2}$Te$_{3}$\xspace}
\newcommand{\bsts}{Bi$_{2-x}$Sb$_{x}$Te$_{3-y}$Se$_{y}$\xspace}
\newcommand{\bstsb}{Bi$_{1.5}$Sb$_{0.5}$Te$_{1.7}$Se$_{1.3}$\xspace}

\newcommand{\bise}{Bi$_{2}$Se$_{3}$\xspace}

\newcommand{\icrn}{$I_{\tiny{\textrm{C}}} R_{\tiny{\textrm{N}}}$\xspace}
\begin{document}
\title{Josephson supercurrent in a topological insulator without a bulk shunt}
\author{M. Snelder$^1$, C.G. Molenaar$^1$, Y. Pan$^2$, D. Wu$^2$\footnote{Current address: School of Physics and Engineering, Sun Yat-Sen University, Guangzhou, China}, Y. K. Huang$^2$, A. de Visser$^2$, A. A. Golubov$^{1,3}$, W.G. van der Wiel$^1$, H. Hilgenkamp$^1$,  M. S. Golden$^2$ and A. Brinkman$^1$}
\address{$^1$ Faculty of Science and Technology and MESA+ Institute for Nanotechnology, University of Twente, 7500 AE Enschede, The Netherlands}
\address{$^2$ Van der Waals - Zeeman Institute, University of Amsterdam, Science Park 904, 1098 XH Amsterdam, the Netherlands}
\address{$^3$ Moscow Institute of Physics and Technology, Dolgoprudny, Moscow 141700, Russia}
\ead{m.snelder@utwente.nl}
\begin{abstract}
A Josephson supercurrent has been induced into the three-dimensional topological insulator \bstsb. We show that the transport in \bstsb exfoliated flakes is dominated by surface states and that the bulk conductivity can be neglected at the temperatures where we study the proximity induced superconductivity. We prepared Josephson junctions with widths in the order of 40 nm and lengths in the order of 50 to 80 nm on several \bstsb flakes and measured down to 30 mK. The Fraunhofer patterns unequivocally reveal that the supercurrent is a Josephson supercurrent. The measured critical currents are reproducibly observed on different devices and upon multiple cooldowns, and the critical current dependence on temperature as well as magnetic field can be well explained by diffusive transport models and geometric effects.
\end{abstract}
%\submitto{\SUST}
\maketitle
\section{Introduction}
Topological insulators (TIs) have conducting surface states with a locking between the electron momentum and its spin \cite{Fu2007, Fu2007a,Hsieh2008,Xia2009,Hsieh2009b,Zhang2009,Hasan2010,Brune2012,Tchakov2013}. Besides bearing promise for high temperature spintronic applications \cite{Hsieh2009a,Jozwiak2013,Li2014}, TIs are also candidate materials to host exotic superconductivity. For example,  $p+ip$ order parameter components \cite{Potter2011,Read2000} and Majorana zero energy states \cite{Nilsson2008,Tanaka2009,Alicea2012,Beenakker2013} have been theoretically predicted. The topological superconductivity can either be intrinsic \cite{Sasaki2011} or proximized by a nearby superconductor \cite{Fu2008,Stanescu2010,Zhang2011}.

The first generation of topological insulators, Bi-based materials such as Bi$_{1-x}$Sb$_x$ alloys, and later \bite and \bise compounds, exhibit topological surface states but also have an additional shunt from the conducting bulk, mainly due to anti-site defects and vacancies \cite{Hsieh2009b,Ren2010,Ren2011}. Josephson junctions \cite{Veldhorst2012,Sacepe2011,Qu2012,Zhang2011,Williams2012, Orlyanchik2013, Cho2013,Wang2012,Oostinga2013,Maier2012,Galletti2014,Sochnikov2013} and SQUIDs \cite{Veldhorst2012a,Qu2012,Kurter2013,Sochnikov2013,Galletti2014} have been realised in these topological surface states, but the practical use of these topological devices is limited by the bulk shunt \cite{Veldhorst2013,Veldhorst2012a}. Secondary and ternary compounds have been engineered to increase the bulk resistance and increase the stability of the surface states. The most promising examples of the latest generation three-dimensional TIs are \bsts \cite{Taskin2011} and strained HgTe\cite{Konig2007, Roth2009, Bouvier2011}.

In this work we report the realization of a Josephson supercurrent across 50 nm of topological insulator \bstsb. We first show that in our \bstsb no bulk conduction is present at low temperatures and that the observed surface states are of a topologically non-trivial nature. We then demonstrate Josephson junction behaviour reproducibly on different flakes and during multiple cooldowns. The width of the superconducting Nb electrodes is very narrow, of the order of 40 nm, anticipating future work on topological devices with only a few modes \cite{Beenakker2011,Beenakker1991,Furusaki1992}.

\section{Transport properties of exfoliated \bstsb flakes}
\bstsb single crystals were obtained by melting stoichiometric amounts of the high purity elements Bi (99.999 \%), Sb (99.9999 \%), Te (99.9999 \%) and Se (99.9995 \%). The raw materials were sealed in an evacuated quartz tube which was vertically placed in the uniform temperature zone of a box furnace to ensure the homogeneity of the batch. The molten material was kept at 850 $^{\circ}$C for 3 days and then cooled down to 520 $^{\circ}$C with a speed of 3 $^{\circ}$ C/h. Next the batch was annealed at 520 $^{\circ}$C for 3 days, followed by cooling to room temperature at a speed of 10 $^{\circ}$C/min\cite{Pan2014}. 

\begin{figure}[t]
\centering 
\includegraphics[width=1\textwidth]{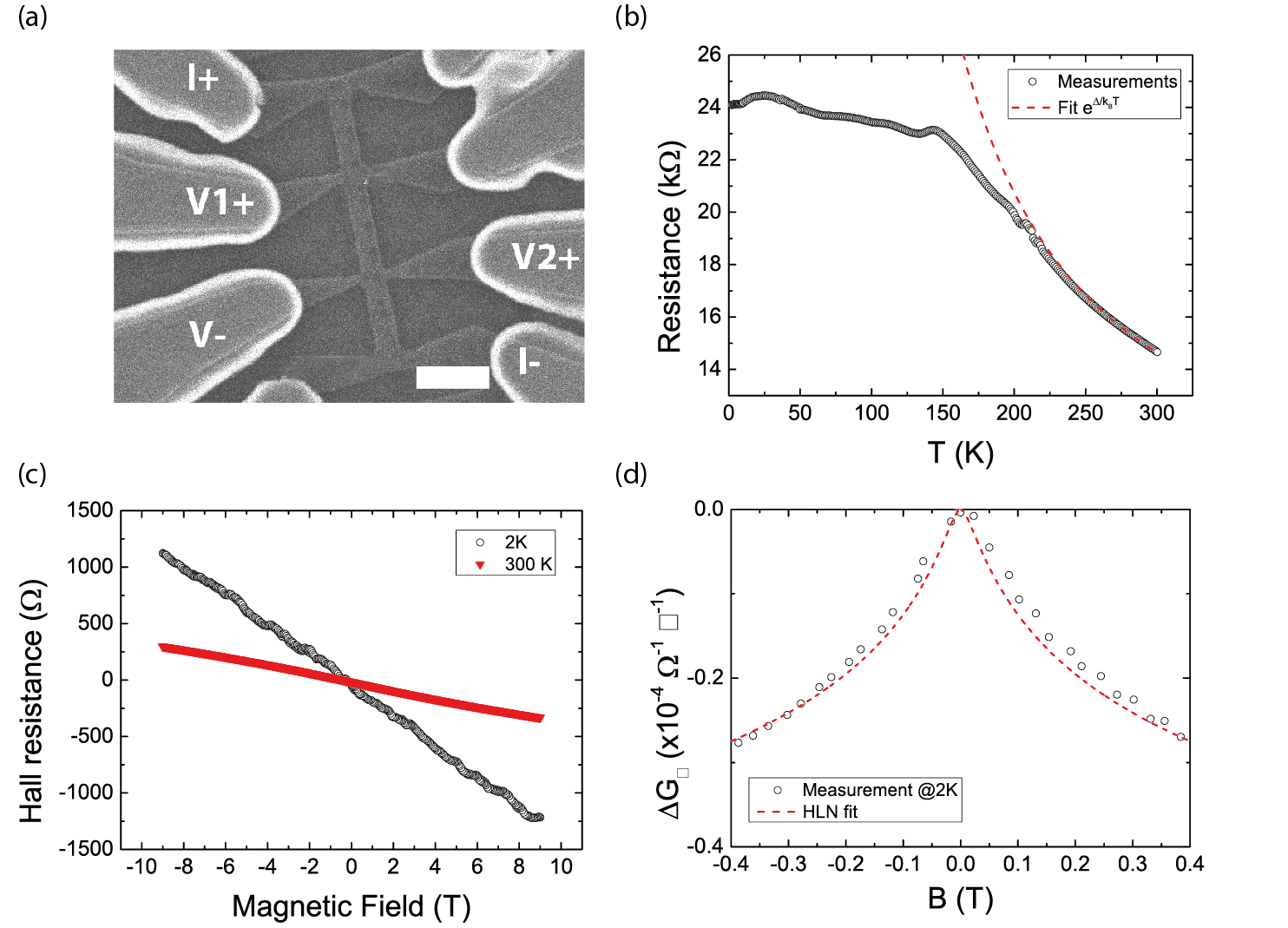}
\caption{(a) Scanning electron microscopy image of a Hall bar of \bstsb. The dimension of this Hall bar is different from the Hall bar of the measurements shown in Fig. 3 (c--d). The Hall bar of the measurements has a width of 560 nm and a length of 6.8 $\mu$m. The dark grey part is etched away. The light gray part between the electrodes is the topological insulator. The white scale bar is 5 $\mu$m. The longitudinal resistance is measured between V1+ and V-, and the Hall resistance between V- and V2+. (b) Typical temperature dependence of the resistance of an exfoliated flake of \bstsb, measured in a Hall bar configuration. The dependence is fitted with a function $e^{\Delta/k_{B}T}$.(c) A typical Hall effect measurement at 2K (black) and 300 K (red). The negative slope implies that the charge carriers are electrons. (d) Measured magnetoconductance, togetherwith a fit of the HLN theory to the data. The best fit is obtained for $\alpha=-1.01$ and $l_{\phi}=144$ nm. } 
\label{fig:1}
\vspace{-15pt} 
\end{figure}

Smooth flakes are prepared using mechanical exfoliation on a silicon-on-insulator substrate. To determine the transport characteristics of \bstsb, Hall bars are prepared using e-beam lithography and argon ion etching on exfoliated flakes with a thickness ranging from 80 till 200 nm. Au electrodes are defined by photolithography and lift-off. During all the fabrication steps the Hall bar is covered with either e-beam resist or photoresist protecting the surface from damage or contamination. The Hall bars are 7 $\mu$m long and 700 nm wide. Figure \ref{fig:1}(a) shows a SEM image of such a Hall bar. 

A typical temperature dependence of the resistance of the Hall bars is shown in Fig. \ref{fig:1}(b).
At high temperature, the crystal exhibits semiconductor-like thermally activated behaviour. Below 150 K the resistance stabilises, indicating metallic surface channels. At high temperatures the transport properties are determined by the bulk of the crystal, while at low temperatures the surfaces provide the dominant charge carriers. To verify this, the high temperature part of the curve is modelled as a semiconductor using $R \propto e^{\Delta/k_{B}T}$. The best fit between 200 K and 300 K  gives $\Delta=$ 18 meV. This means that the Fermi energy is positioned 18 meV below the bottom of the conduction band. The entire bulk band gap would be larger than 18 meV. At 300 K, the bulk contribution is dominant and allows for a one-band interpretation of the Hall effect measurement (see Fig. \ref{fig:1}(c)) at this temperature, giving a bulk carrier density of 10$^{17}$ cm$^{-3}$. Extrapolating the carrier freeze out to low temperatures we find a negligible bulk conduction at low temperatures. At low temperatures all transport is due to the surface states. Hall and longitudinal resistance measurements at 2 K are used to determine the electron density and mobility of the surface states. We reproducibly obtain surface electron densities in the range of 10$^{12}$ - 10$^{13}$ cm$^{-2}$ with mobilities between 120 and 450 cm$^{2}$/Vs. The resulting mean free path of the order of 10 to 40 nm is comparable to the mean free path of the electron-like surface states found by Taskin \textit{et al.} \cite{Taskin2011}. 

Figures \ref{fig:1}(d) shows the change in longitudinal conductance as function of applied perpendicular magnetic field. To verify the topological character of the surface states the Hikami-Larkin-Nagaoka theory \cite{Hikami1980} is used to fit the low-field magnetoconductance in perpendicular field,
\begin{eqnarray}
\Delta G_{\square} (B_{\perp})-\Delta G_{\square}(0)=\alpha \frac{e^{2}}{2\pi^{2}\hbar}\left[\psi\left(\frac{1}{2}+\frac{\hbar }{4el_{\phi}^{2}B_{\perp}}\right)-\mbox{ln}\left(\frac{\hbar } {4el_{\phi}^{2}B_{\perp}}\right)\right],
\end{eqnarray}
where $G_{\square}=\frac{L}{WR}$ with $L$ being the length and $W$ the width of the Hall bar, $R$ is the measured resistance, $\psi$ is the digamma function, $l_{\phi}$ is the inelastic scattering length, $\alpha$ is a parameter indicating the strength of the spin-orbit interaction and $B_{\perp}$ is the perpendicular magnetic field. If the spin-orbit interaction is weak, a positive value of $\alpha=1$ is expected, as opposed to strong spin-orbit interaction where a negative value of -0.5 is expected\cite{Hikami1980}. Due to the chiral-spin texture of a topological insulator and the contribution of both top and bottom surfaces in the transport measurements, an $\alpha$ parameter of $-1$ is expected. In Fig. \ref{fig:1}(d) the magnetoconductance is fitted with $l_{\phi}=$ 144 nm and $\alpha=-$1.01 which would be in good agreement with the presence of just the bottom and top topologically non-trivial surface states.

However, we cannot rule out the presence of additional trivial two-dimensional metallic surface states that could originate from surface band bending. The possibility of the conduction band bending down below the Fermi energy has been observed in \bstsb \cite{Taskin2011}. Angle resolved photo-electron spectroscopy has revealed that the presence of these non-topological surface states is depending on surface absorbents and could vary over time\cite{Golden2014}. 

\section{Junction fabrication}
\begin{figure}[t]
\centering 
\includegraphics[width=1\textwidth]{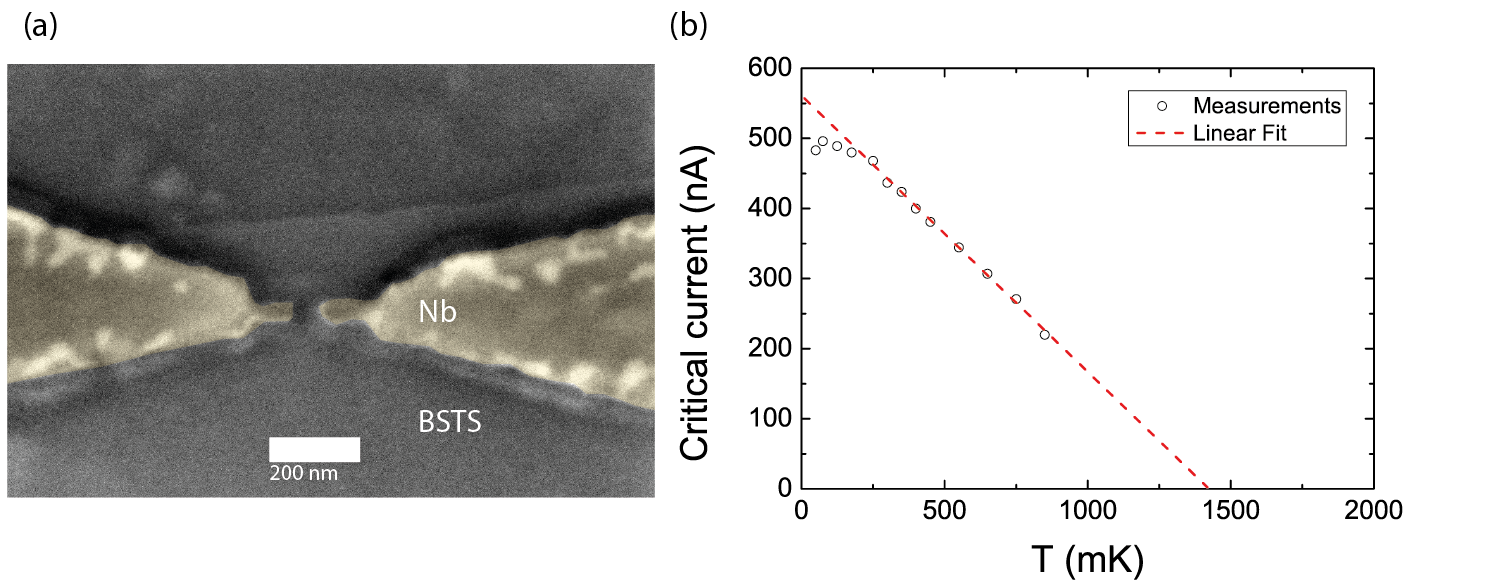}
\caption{(a) Scanning electron microscopy image of a Nb/\bstsb/Nb junction. The Nb layer has been coloured yellow for clarity. The junction is 57 nm long and 38 nm wide. (b) The dependence of the critical current of the Nb nanowire as function of temperature. From the linear fit to the data point in the range of 300-850 mK, the critical temperature is extrapolated to be 1.4 K. Close to the transition temperature, the gap and therefore the critical current can be approached by a linear dependence \cite{Ferrell1964}. 
} 
\label{fig:2}
\vspace{-15pt} 
\end{figure}

Now that we have verified the topological nature of our crystals we turn to fabricating Josephson junctions with \bstsb as the weak link. Transferred similarly to the devices for determining the transport properties, smooth flakes on Si/SiO$_{2}$ substrates are used for devices. E-beam lithography is used to define the junctions and the contact pads. Thereafter we perform a 30 second low voltage etching step to avoid large damage to the surface followed by sputtering in-situ 25 nm Nb and 2.5 nm Pd to protect the Nb layer, and lift-off of the excess material. The resulting Nb/\bstsb/Nb junction has been covered by photoresist during the entire process and no damage from etching or growth has occurred to the crystal surface between the electrodes. In figure \ref{fig:2}(a) a junction is visible in a SEM image, and similar junctions have been prepared on different flakes. 
The width is about 40 nm and the electrode separation is about 50 nm. The motivation for the junction length is based on our earlier work on Bi$_{2}$Te$_{3}$ with junctions ranging from 50-250 nm \cite{Veldhorst2012}. The mean free path in \bstsb is ten times smaller compared to \bite which implies that the junction length should also be reduced by a factor of ten to realise sufficient coupling between the electrodes. Simultaneously, this design takes a step forward towards experiments for the observation of a quantized supercurrent where only a few modes should be present\cite{Beenakker2011,Beenakker1991,Furusaki1992}. In order to achieve the presence of a few modes, the junction width should be of the order of the Fermi wave length. For a 40 nm wide junction and a Dirac velocity of $4.5 \times 10^5$ m/s this means that the Fermi level would have to be tuned within 50 meV from the Dirac point by means of gating, which is quite reasonable. To realise these dimensions a thin layer of PMMA (80 nm) is used which limits the thickness of Nb that could be grown. The reduction of the dimensions of the Nb leads reduces the gap of Nb. A 40 nm wide and 250 nm long Nb nanowire is prepared in the same run to determine the properties of the niobium with these dimensions. In figure \ref{fig:2}b the critical current of this nanowire as a function of temperature is presented, and a reduction of the critical temperature from 9 K of the Nb of our thin film process to about 1.4 K is visible.

\section{Results}
Measurements are performed in a cryogen free dilution refrigerator with low pass filtering of current and voltage signals using pi filters, printed circuit board copper powder filters \cite{Mueller2013}, and RC filters. We measured the critical current as function of temperature and the modulation of the critical current by an applied magnetic field at 30 mK. We will start this section with a discussion of the temperature dependence followed by an analysis of the Fraunhofer pattern.
\subsection{Temperature dependence of the critical current}
\begin{figure}[t]
\centering 
\includegraphics[width=\textwidth]{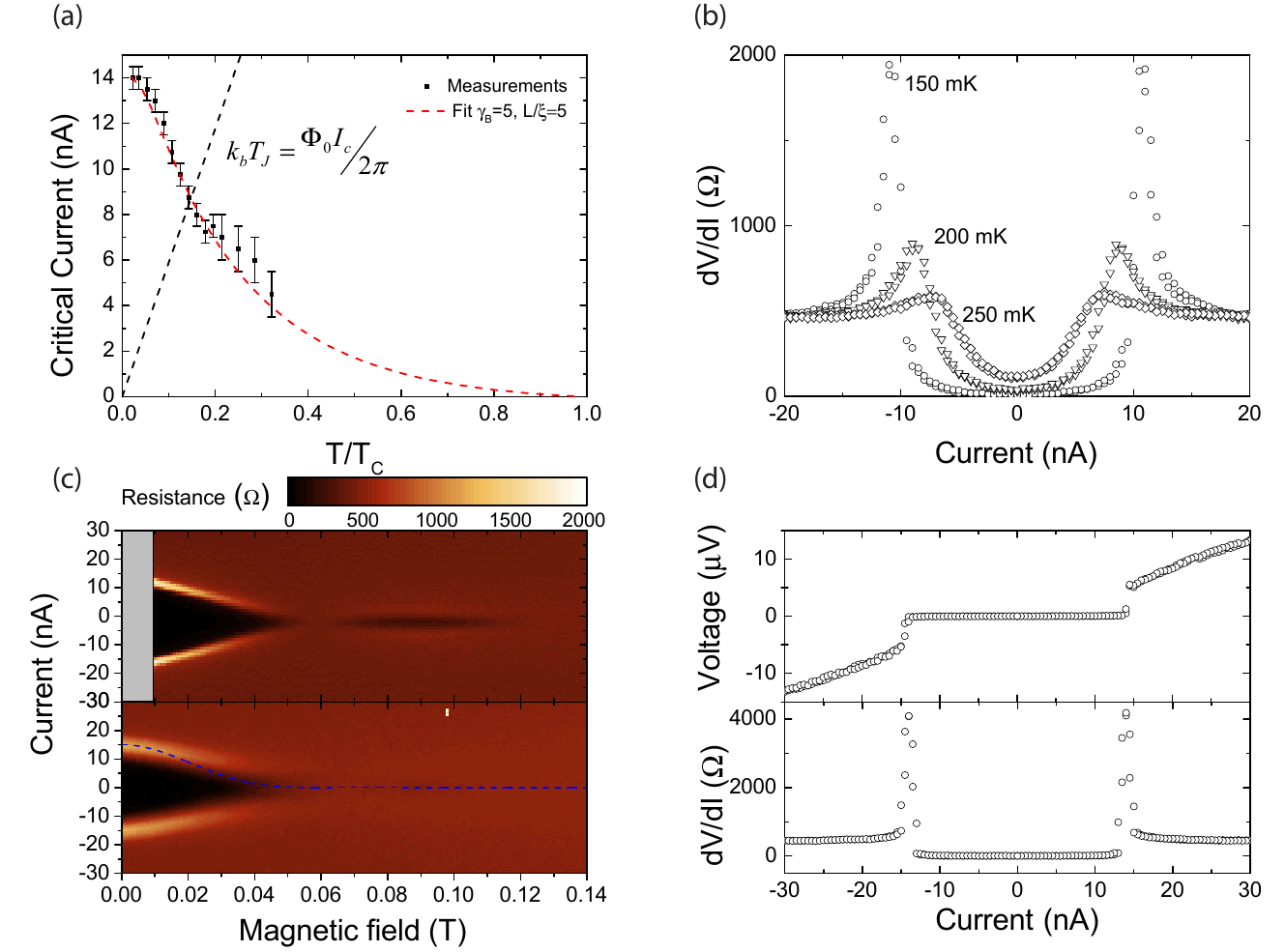}
\caption{Josephson characteristics of the Nb/\bstsb/Nb junction illustrated in Figure \ref{fig:2}.
(a)~Temperature dependence of the critical current. Above the thermal limit, indicated by the black dashed line, the critical current is determined by the resistance peaks. Below the thermal limit the width of the resistance dip is used as estimate. The red dashed line is a fit for diffusive junctions with the Usadel equation, see main text.
(b)~Derivative of the current-voltage characteristic above, at, and below the thermal limit. Below the thermal limit the resistance is zero and coherence peaks are visible. These peaks disappear above the thermal limit of $~$200mK and only a resistance valley remains.
(c)~Modulation of the critical current by an applied perpendicular magnetic field. The upper and lower frames are measurements during different cooldown cycles. The upper frame was the first cooldown using Al bond wires which have reduced cooling power below 10 mT, resulting in a sharp step in the critical current at this point. The bottom frame uses Au bond wires. This second cooldown exhibited a slightly reduced critical current and ergo reduced visibility of the first lobe of the diffraction pattern. The dotted line is a model of the critical current by Barzykin \textit{et al.} \cite{Barzykin1999} using the junction parameters found in (a).
(d)~Current-voltage characteristic and derivative at base temperature. The critical current of 14 nA and normal state resistance of 460 $\Omega$ result in an \icrn product of 6.4 $\mu$V, consistent with a junction with high $\gamma_B$ and $L \geq \xi_N$. } 
\label{fig:CriticalCurrent}
\vspace{-15pt} 
\end{figure}

At 30 mK we observe a critical current of 14 nA for a junction with 57 nm electrode separation, visible in figure \ref{fig:2}(a), and 4.8 nA for a 80 nm long junction. Using the extracted critical temperature of the Nb nanowire as the transition temperature for the leads, the normalised temperature dependence of the critical current is plotted in figure \ref{fig:CriticalCurrent}(a) for the 57 nm junction. The boundary between Josephson coupling and thermal noise $(\hbar I_{0}/e)/(k_{B}T)=1$ is illustrated with the dashed line. In figure \ref{fig:CriticalCurrent}(b) the dV/dI characteristics around the thermal limit reveal indeed that a zero resistance state and coherence peaks are only visible below the thermal limit. The coherence peaks are used for determining the critical current below the thermal limit. Above the thermal limit the width of the resistance valley is used as an estimate.

The nature of weak link is determined by fitting the temperature dependence of the supercurrent. For diffusive SNS junctions, i.e. junctions where the junction length $L$ is larger than the electron mean free path, the Usadel equation is used to describe the temperature dependence \cite{Usadel1970,Zaitsev1984,Kupriyanov1988},
\begin{equation}
J=\frac{2\pi k_{B}T}{e\rho_{N}}\mbox{Im} \sum _{\omega_{n}>0}\frac{G_{N}^{2}}{\omega_{n}^{2}}\Phi_{N}\frac{d}{dx}\Phi_{N},
\end{equation} where $\rho_{N}$ is the resistivity of the N layer, $\Phi_{N}$ is the self-consistently determinded induced order parameter function in the N layer with $G_{N}$ the corresponding normal Green function and $\omega_{n}=\pi k_B T (2n+1)$ the Matsubara frequencies, where $n \geq 0$ is integer. As there is no analytical expression for arbitrary length and barrier transparency, the expression was solved numerically with three fitting parameters\cite{Golubov1995}, $\gamma=\frac{\rho_{s}\xi_{s}}{\rho_{N}\xi_{N}}$, $\gamma_{B}=\frac{R_{B}}{\rho_{N}\xi_{N}}$ and $\xi_{N}$. Here, $\gamma$ is the ratio between the resistivities and coherence lengths in the superconducting leads and the topological insulator surface state. The resistivity of Nb is much smaller than that of \bstsb which gives $\gamma \ll1$. $\gamma_B$ is proportional to the interface resistance per unit area $R_{B}$ between the S and N layers. Finally, $\xi_N$ is the coherence length in the normal region at $T_{C}$. For the junction illustrated in figures \ref{fig:2}(a) and \ref{fig:CriticalCurrent}, the best fit is obtained for $\gamma_{B}=5$ and $\xi_{N}=11.4$ nm. The other junctions give values of the same order of magnitude. The large $\gamma_{B}$ is in good agreement with the low excess current in the $IV$ curves which implies a large barrier or low transparency at the interface. Taking the $T_{c}$ of the Nb wire, we get for $\xi_{N}\left( T_{c}\right)=\sqrt{\frac{\hbar D}{2 \pi k_{B}T_{c}}}$ a value of 28 nm which is also in good agreement with the fitting parameters.
We note, that the junctions have a small \icrn product, $\sim 7 \mu$V. We argued that $\gamma \ll 1$. Together with the fitted values of $\gamma_{B}$ and $\xi_{N}$ the \icrn product is either of the form $\propto V_{0}/\gamma_{B}$ or \icrn$\propto V_{0}e^{-L/\xi_{N}}$ \cite{Golubov2004}. Substitution of $\gamma_{B}$, $\xi_{N}$ and the junction length $L$ we estimate from this relation that the \icrn should be in the order of 1-10 $\mu$V which is in good agreement with the measured product value. 
\subsection{Critical current as a function of magnetic field}
In Fig. \ref{fig:CriticalCurrent}(c) measurements during two separate cooldowns of the I$_{\tiny{\textrm{C}}}$(B) pattern at 30 mK are shown for the same junction. The changes in the supercurrent in the second cooldown indicate a slight evolution in time. The upper frame shows the Fraunhofer pattern of the first measurement. Due to the use of Al bond wires, the cooling of the sample is reduced below the critical field of Al. These measurements are greyed out because the temperature increased in that region, we estimate by 100 mK. A second cooldown using Au bonds results in the lower panel Fraunhofer pattern. Here, the critical current has reduced slightly resulting in a decreased visibility of the lobes of the Fraunhofer pattern.
The Fraunhofer patterns are fitted by the critical current derived from the Usadel equations for arbitrary $W$ and $L$ with open boundaries, as described by Barzykin \textit{et al.} \cite{Barzykin1999},
\begin{eqnarray}
I_{c} &\sim  \sum^{\infty}_{l=-\infty}(-1)^{l}S_{l}(L/2)S'_{l}(L/2) \left(\frac{\sin \pi (\nu+l)/2}{\pi(\nu+l)/2)}-(-1)^{l}\frac{\sin \pi (\nu-l)/2}{\pi(\nu-l)/2)}\right)^{2},\nonumber \\
S_{l}(u) &= \sqrt{|u|/2\pi}\left(q_{T}^{2}+\pi^{2}l^{2}/W^{2}\right)^{1/4} K_{1/2}\left(\sqrt{u^{2}\left(q_{T}^{2}+\pi^{2}l^{2}/W^{2}\right)}\right),\nonumber \\
S_{l}'(u) &= \frac{d}{du}S_{l}(u),\label{FPmodel}
\end{eqnarray} where $\nu=\phi/\phi_{0}$ with $\phi_{0}=\hbar/2e$ is the normalised flux, $q_{T}=1/\xi_{N}^{2}$ and $K_{1/2}$ is a modified Bessel function of the second kind. For our junction dimensions it follows from this model that the first period is tripled, i.e. the first minimum is at $\Phi=3\Phi_{0}$ and the following minima are separated by $2\Phi_{0}$ intervals. The doubling of the period is not restricted to diffusive junctions and was, in fact, also predicted \cite{Barzykin1999} and observed \cite{Heida1998} in ballistic junctions for width and length ratios in the order of one. 
The first minimum in the measurements occurs at a field value of 0.07 T. The width of the junction is 37 nm. For a flux of 3$\phi_{0}=6.2\cdot 10^{-15}$Wb this implies that the junction effective length (including the penetration depths) is about 2 $\mu$m. The obtained London penetration depth is then about 1 $\mu$m. This large London penetration depth compared to bulk Nb (47 nm)\cite{Maxfield1965} can be explained by the reduced dimensions of the junctions. The increase of the London penetration depth with decreasing film thickness is studied in Ref. \citeonline{Gubin2005}. A film thickness of 25 nm gives a London penetration depth above 100 nm and a decrease of the critical current to 8 K. Due to the reduction of width of the junction in our devices, we end up with a $T_{C}$ of 1.4 K which consistently implies an even larger increase of the London penetration depth. 
The coherence length found from the fitting of the critical current together with the width and the length of the junction and London penetration depth serve as inputs for the model described in equation \ref{FPmodel}. The theoretical expectation is shown in figure \ref{fig:CriticalCurrent}(c). The relative small critical current of the higher order peaks in the data  is found to be well explained by the model.
\section{Discussion}
We showed that the transport in the \bstsb flakes is dominated by surface states at low temperatures where we study proximity induced superconductivity. We prepared Josephson junctions with widths in the order of 40 nm and lengths in the order of 50 to 80 nm on several \bstsb flakes and measured them during several cooldowns to 30 mK. The Fraunhofer patterns unequivocally reveal that the supercurrent is a Josephson supercurrent. The measured critical currents are reproducibly observed on different devices and upon multiple cooldowns, and the measurement can be well explained by diffusive transport models and geometric effects. The predicted $4\pi$ periodic Josephson effect can only be observed in the future for similar devices with just a single perpendicular mode \cite{Snelder2013} and when measured faster than the relaxation rate of a quasiparticle inside an Andreev bound state.

The realization of a Josephson supercurrent in junctions with dimensions in the order of tens of nanometers on a topological insulator dominated by surface states at low temperatures, is an important technological step towards advanced devices necessary for the observation of a quantized supercurrent and confirming the presence of a Majorana bound states in such devices \cite{Fu2008}. Future work will focus on top and bottom gating of the surface states to eliminate potential additional trivial surface states from band bending.  
\ack
We would like to thank Frank Roesthuis and Dick Veldhuis for support during fabrication and Elia Strambini for support during the measurements. We acknowledge Martin Stehno for useful discussions. This work is supported by the Netherlands Organization for Scientific Research (NWO), by the Dutch Foundation for Fundamental Research on Matter (FOM), by the European Research Council (ERC) and supported in part by Ministry of Education and Science of the Russian Federation, grant no. 14Y26.31.0007. 
\section*{References}
\bibliographystyle{iopart-num-nourl}
\bibliography{bsts}
\end{document}